\documentclass[12pt,leqno]{article}

\usepackage{graphicx}
 \font\gotb eufm10 scaled \magstep1
\newcommand{\bb}{\bibitem}
\newcommand{\cc}{\cite}
\newcommand{\vp}{\varphi}
\newcommand{\sss}{\sigma}
\newcommand{\al}{\alpha}
\newcommand{\Om}{\Omega}

\newcommand{\lt}{\left}
\newcommand{\rt}{\right}

\newcommand{\ra}{\rangle}

\newcommand{\F}{{\cal F}}

\newcommand{\s}{\hat S}

\newcommand{\A}{\hat A}

\newcommand{\AAA}{\hbox{\gotb A}}

\newcommand{\HHH}{\hbox{\gotb H}}
\newcommand{\QQ}{\hbox{\gotb Q}}
\newcommand {\QQQ}{\hbox {\gotb Q}_{\xi}}
\newcommand {\qqq}{\hbox {\gotb Q}_{\xi'}}
\newcommand {\vx}{\vp_{\xi}}

\newcommand{\BB}{\hbox{\gotb B}}
\newcommand{\bea}{\begin{eqnarray} \label}
\newcommand{\eeq}{\end{equation}}
\newcommand{\beq}{\begin{equation} \label}
\newcommand{\eea}{\end{eqnarray}}
\newcommand{\nn}{\\ \nonumber}
\newcommand{\rr}[1]{(\ref{#1})}

\newcommand{\post}{{\sc Postulate}}

 \author{D.A.Slavnov}

\title{QUANTUM TELEPORTATION\thanks{Theoretical and Mathematical Physics, 157(1): 1433-1447 (2008) }}

   \date{}

\begin{document}

  \maketitle

 \begin{abstract}

In the framework of an algebraic approach, we consider a quantum
teleportation procedure. It turns out that using the quantum
measurement nonlocality hypothesis is unnecessary for describing
this procedure. We study the question of what material objects are
information carriers for quantum teleportation.
\end{abstract}

\section{Introduction}
The term "teleportation" in the title of this paper can provoke
totally justified suspicion in a reader of a scientific journal.
This term is so closely associated with science fiction and
pseudoscientific literature. Nevertheless, this term is widely
used in the scientific field often called "quantum information
physics" (see, e.g., \cc{pqi}). This field has developed
intensively in recent years, and great hopes of very interesting
practical applications are connected with it.

A specific method for transferring information is called
"teleportation." This method has some mysterious features even in
the purely scientific literature because the material information
carrier is not clearly indicated. Instead, nonlocality, which is
supposedly inherent in quantum measurements, is cited.

Here, we try to give a grounded, visual picture of teleportation,
indicating what material carrier transfers information of one or
another type. We use a special version of the algebraic approach
to quantum theory. A detailed description of this approach can be
found in  \cc{slav1}, and a condensed one, in  \cc{slav2}. In
those papers, we used an inductive method to construct the theory.
First, we studied physical phenomena, then noted the physical
patterns inherent in them, and finally described these patterns in
the form of mathematical axioms. Here, we assume that the reader
already has a knowledge of the phenomenological justification of
the axioms and restrict ourself to formulating them. Also, we do
not describe how the standard mathematical apparatus of quantum
mechanics can be obtained using these axioms. We again refer the
interested reader to \cc{slav1,slav2}.

\section{Main statements of the approach}

The central notion used in this approach is an "observable." An
observable is an attribute of a physical system whose numerical
value can be obtained using a certain measuring procedure. All the
observables are assumed to be dimensionless. All the measurements
are divided into reproducible and nonreproducible ones and also
into compatible and incompatible ones. Compatible measurements are
conducted using compatible measuring devices. If there exist
compatible measuring devices for a group of observables, then such
observables are said to be compatible (simultaneously measurable).

\

\post{} 1. Observables $\A$ of a physical system are Hermitian
elements of some   $C^*$-algebra~\cc{dix}~\AAA, ($\A\in \AAA,
\A^*=\A$).

\

We let ~$\AAA_+$ ($\AAA_+\subset \AAA$) denote the set of
observables.

\

\post{} 2. The set of compatible observables is a maximal real
associative commutative subalgebra ~$\QQQ$ of the algebra~\AAA
\quad ($\QQQ\subset\AAA_+$).

\

The index~$\xi$ ranging a set  $\Xi$ distinguishes one such
subalgebra from another. There is only one such subalgebra for a
classical system; there are infinitely many for a quantum system
\cc{slav1}.

We regard the set  $\AAA_+$ as a mathematical representation of
the physical system under study and the sets $\QQQ$ as
mathematical representations of the corresponding classical
subsystems of the physical system. These classical subsystems are
open (interacting between themselves) and do not have their own
dynamics. The state of a classical system is its attribute that
uniquely predetermines the results of measurements of all the
observables of the system. Therefore, we formulate the following
postulate.

\

\post{} 3. The state of a classical subsystem whose observables
are elements of the subalgebra  $\QQQ$ is described by a character
of this subalgebra.

\

We recall that a homomorphic map of the associative commutative
algebra to the set of  numbers is called the character
$\vx(\cdot)$ of this algebra:
$\A\stackrel{\vx}{\longrightarrow}\vx(\A)$, $\A\in \QQQ$ (see,
e.g.,  \cc{dix}).

Because the observables belonging to the subalgebra  $\QQQ$ are
compatible, there exists a set of measuring devices designed for
compatible measurements of these observables. We say that these
devices belong to the $\xi$-type.

The set  $\AAA_+$ of observables of a quantum system can be
regarded as a collection of subsets  $\QQQ$. Therefore, the
quantum system can be regarded as a set of corresponding open
classical subsystems. Each observable of the quantum system
belongs to a certain subset $\QQQ$. Accordingly, if the states of
all classical subsystems were known, then we could predict the
result of measuring any observable of the quantum system. Based on
this, we call the set $\vp=[\vx] \quad (\xi\in\Xi)$ of functionals
$\vx(\cdot)$ each of which is the character of the corresponding
subalgebra~$\QQQ$, the elementary state of a physical system. The
following postulate is central in the described approach.

\

\post{} 4. The result of each individual measurement of the
observables of a physical system is determined by the elementary
state of this system.

\

This postulate does not say that the result is uniquely
determined. The point is that the same observable A can
simultaneously belong to several subalgebras  $\QQQ$. Therefore,
we can use devices of different types to measure it. We are
accustomed to the fact that if the device is "good," then the
measurement result is independent of the device. But this holds
only in the case where all the devices can be calibrated in the
same way. Such a calibration is, in principle, possible in the
classical case. But such a calibration, as shown in  \cc{slav1},
cannot be implemented for quantum systems because of the presence
of incompatible measurements. Therefore, measurement results in
the quantum case depend on two factors: the elementary state and
the type of device used for measuring. In the general case, we say
that measuring devices belong to the $\xi$-type if  $\A$ is
obtained as a result of measurements of the observable  $\vx(\A)$.

In some cases, measurement results can be independent of the type
of measuring device, i.e.,  $  \vx(\A)=\vp_{\xi'}(\A) $ for all
$\QQQ,\;\qqq,$ containing $\A$. In this case, we say that the
elementary state $\vp$  is stable for the observable  $\A$.

We note that in the majority of proofs demonstrating that the
existence of local physical reality determining measurement
results is impossible, it is tacitly assumed that this result is
independent of the type of measuring device (see, e.g.,  \cc{ksp,
ghz}).

There is one more hindrance to the unique prediction of
measurement results in the quantum case. The point is that it is
impossible to determine the elementary state of a system uniquely
in experiments. Only compatible measuring devices can be used to
fix it. Using such devices, we can determine the functional
$\vx(\cdot)$ only for one value of  $\xi$ ($\xi=\eta$). All the
other functionals  $\vx(\cdot)$ contained in the elementary state
$[\vx]$, remain undetermined. Figuratively speaking, we can say
that the elementary state is a holographic image of a physical
system. Using classical measuring devices, we can find only a
plane image. In this case, each measurement changes an original
holographic pattern. Therefore, it is impossible to obtain a
complete holographic image.

We unite all the elementary states  $[\vx]$ having the same
restriction to the subalgebra  $\QQ_{\eta}$, i.e., the same
functional $\vp_{\eta}$, into the equivalence class
$\{\vp\}_{\eta}$. Thus, it is possible in experiments to uniquely
fix only the equivalence class to which the elementary state of
interest belongs. If we know that some elementary state
$\vp=[\vx]$ belongs to the equivalence class  $\{\vp\}_{\eta}$,
then we can uniquely predict what result can be obtained in the
measurement of the observable $\A\in\QQ_{\eta}$ this result is
$\vp_{\eta}(\A)$. But if  $\A\notin\QQ_{\eta}$, then it is
impossible to say anything definite about the measurement result.
For different elementary states belonging to  $\{\vp\}_{\eta}$,
the measurement results are different. The quantum state fixed by
certain values of the observable $\A$ from the subalgebra   has
such physical properties in the standard quantum mechanics
$\QQ_{\eta}$.

For the observables  $\A\notin\QQ_{\eta}$ only the probabilities
of the determination of different results can be predicted. It is
postulated in the standard mathematical apparatus of quantum
mechanics that these probabilities are determined by either a
vector of some Hilbert space or a statistical operator (the
density matrix). This postulate holds in practice, but it is
impossible to understand its justification at the intuitive level.
Using the approach given here, we can justify this postulate. It
is natural to assume that the probability distribution of
measurement results is determined by the probability distribution
of different elementary states in the equivalence class
$\{\vp\}_{\eta}$. Therefore, we formulate the following postulate.

\

\post{} 5. The equivalence class  $\{\vp\}_{\eta}$ can be endowed
with the structure of a probability space.

\

We note that an elementary state satisfies the requirements that
the classical (Kolmogorov \cc{kol}) probability theory imposes on
an elementary event: one and only one elementary event occurs in
each trial, i.e., elementary events exclude each other. This
allows endowing $\{\vp\}_{\eta}$ with the standard (classical)
structure of a probability space $(\Om,\F, P)$ (see, e.g.,
\cc{kol,nev}). Here,  $\Om$ is the set of elementary events
(states), $\F$ is the Boolean $\sss$-algebra of events $F$, $P$ is
the probability that an event  $F\in\F$ occurs. The event  $F$ is
a subset of the set  $\Om$. It is assumed that the event  $F$,
occurs in a trial if one of the elementary events belonging to
this subset occurs.

Thus, constructing a special artificial quantum probability theory
for quantum systems is unnecessary; indeed, the standard
probability theory can be used. It is only necessary to take into
account that not all possible Boolean algebras $\F$ are admissible
in the quantum case   \cc{slav1, slav2}. If this feature of
quantum systems is not taken into account, then it is easy to
obtain the Bell inequalities \cc{bel1,bel2,chsh}, which contradict
experimental data.

If $\{\vp\}_{\eta}$ is endowed with the structure of a probability
space, then we can use standard probability theory methods to
easily construct \cc{slav1} the functional  $\Psi_{\eta}(\A)$,
that specifies the mean of an observable $\A$ in the equivalence
class  $\{\vp\}_{\eta}$. To identify this functional with the
quantum mean of the observable  $\A$, this functional must be
linear. Therefore, we must formulate the following postulate.

\

\post{} 6. The probability structure of the equivalence class
$\{\vp\}_{\eta}$ is such that the functional  $\Psi_{\eta}(\A)$ is
linear on the algebra \AAA.

\

In this case, we can assume that the functional  $\Psi_{\eta}(\A)$
describes the corresponding quantum state. In other words, we can
formulate the following definition in this approach.

\

{\sc Definition}. We call the class $\{\vp\}_{\eta}$
$\vp_{\eta}$-equivalent elementary states that are stable on the
subalgebra  $\QQ_{\eta}$ the quantum state $\Psi_{\eta}(\A)$. This
equivalence class must admit the described structure of a
probability space.

   \

The quantum state thus defined is pure \cc{slav1}.

We have previously noted that the result of a particular
measurement of an observable can depend on the type of measuring
device used to measure it. At the same time, experiments show that
the following assertion holds.

\

\post{} 7. The mean of an observable in a fixed quantum state is
independent of the type of measuring device used to determine it.

\

Having the  $C^*$-algebra $\AAA$ and the linear functional
$\Psi(\cdot)$ on this algebra and using the canonical
Gelfand-Naimark-Segal construction, we can realize the
representation of this algebra (see, e.g., \cc{emch}) by bounded
linear operators in a Hilbert space $\HHH$:

 $$ \A\leftrightarrow\Pi(\A), \quad \A\in\AAA,
\quad \Pi(\A)\in\BB(\HHH), $$ where $\BB(\HHH)$ is the set of
bounded linear operators in  $\HHH$. In this case, the mean
$\langle\A\rangle$ of the observable  $\A$ in the quantum state
$\Psi $ can be expressed as the mathematical expectation of the
operator  $\Pi(\A)$:
 $$
 \langle\A\rangle=\langle\Psi|\Pi(\A)|\Psi\rangle,
 $$
where $|\Psi\rangle\in\HHH$ is the corresponding vector of the
Hilbert space.

We call the set of physical systems whose elementary states form
the equivalence class  $\{\vp\}_{\eta}$ the quantum ensemble.
Thus, the standard mathematical apparatus of quantum mechanics can
be used to describe quantum ensembles but not an individual
quantum system. An elementary state is the adequate mathematical
characteristic of an individual quantum system.

The same elementary state can be considered as an element of
different sets. The probability characteristics that are
associated with a particular physical system significantly depend
on these sets. Therefore, in the proposed approach, we must
reformulate the assertion "the physical system under study is in a
given quantum state" (this assertion is usual in the standard
approach) as "the physical system under consideration is in an
elementary state that belongs to a given quantum state." We can
assume that the same elementary state can simultaneously belong to
another quantum state. Thus, a quantum state is not quite an
objective characteristic of an individual physical system.

\section{Entangled states}

So-called entangled states play the central role in the quantum
teleportation procedure. In the literature, entangled states
typical of a two-particle system in which each of the particles
can be in two orthogonal quantum states $|\pm\ra$ are most often
considered:
  \bea{3}
|\Psi^{(-)}\ra_{12}&=&\frac{1}{\sqrt{2}}\lt[|+\ra_1|-\ra_2-
|-\ra_1|+\ra_2\rt],\nn
|\Psi^{(+)}\ra_{12}&=&\frac{1}{\sqrt{2}}\lt[|+\ra_1|-\ra_2+
|-\ra_1|+\ra_2\rt],\nn
|\Phi^{(-)}\ra_{12}&=&\frac{1}{\sqrt{2}}\lt[|+\ra_1|+\ra_2-
|-\ra_1|-\ra_2\rt],\nn
|\Phi^{(+)}\ra_{12}&=&\frac{1}{\sqrt{2}}\lt[|+\ra_1|+\ra_2+
|-\ra_1|-\ra_2\rt].
 \eea
These states are often called the Bell states. Their
characteristic feature is as follows. Any of these states can be
selected using measuring devices, but after such a selection, it
is impossible to uniquely predict in which of the two possible
quantum states  $|+\ra$ or $|-\ra$ we find each of the particles
in the subsequent measurement. At the same time, after the quantum
state of one of the particles is measured, the state of the other
particle is uniquely predicted.

The state  $|\Psi^{(-)}\ra_{12}$ is usually considered in the
discussion of the Einstein-Podolsky-Rosen paradox~\cc{epr} and is
therefore often called the EPR state. Thus, the system consisting
of two particles with spin 1/2 was considered in the version
proposed by Bohm~\cc{bom}. Then $|+\ra$ is the quantum state with
the spin projection on the $z$ axis equal to  $+1/2$, and $|-\ra$
is the state with the projection equal to  $-1/2$.

In the state  $|\Psi^{(-)}\ra_{12}$ the total spin  ${\bf S}={\bf
S_1}+ {\bf S_2}$ is zero. The state  $|\Psi^{(+)}\ra_{12}$ is the
state with total spin 1 and the projection of the total spin on
the  $z$, axis equal to 0,
$\frac{1}{\sqrt{2}}\lt(\Phi^{(+)}\ra_{12}+|\Phi^{(-)}\ra_{12}\rt)$
is the state with total spin 1 and the projection  $+1$,  and the
state
$\frac{1}{\sqrt{2}}\lt(\Phi^{(+)}\ra_{12}-|\Phi^{(-)}\ra_{12}\rt)$
is the state with total spin 1 and the projection  $-1$. It is
also convenient to use a similar terminology for other two-level
systems, assuming that the states  $|\Psi^{(\pm)}\ra_{12}$,
$|\Phi^{(\pm)}\ra_{12}$ are characterized by the corresponding
values of observables, which we call quasispins.

The characteristic feature of the state  $|\Psi^{(-)}\ra_{12}$ is
its spherical symmetry. Therefore, it preserves its form if the
projections on the $x$ axis or, generally, on an arbitrary
direction n are considered instead of the projections on the  $z$
axis. In this state, for any ${\bf n}$, the relation
$$S_{{\bf
n}1}+S_{{\bf n}2}=0 $$ holds, where  $S_{{\bf n}1}(S_{{\bf n}2})$
is the (quasi)spin projection of the $i$-th particle on the
direction ${\bf n}$. For such a system, the EPR paradox consists
in the following. According to the standard ideology of quantum
mechanics, none of the particles in the state
$|\Psi^{(-)}\ra_{12}$ has a definite value of the (quasi)spin
projection on the direction ${\bf n}$. That the first particle
takes a certain value of the projection as a result of the action
of the measuring device seems quite probable. But it is very
difficult to understand how the measuring device could affect the
second particle if this particle and this device are located in
regions separated by a spacelike interval.

In the framework of the scheme described in the preceding section,
the EPR paradox is trivially solved. The quantum state
$|\Psi^{(-)}\ra_{12}$ describes not one two-particle system but an
ensemble of such systems. Let the index $\al$ distinguish one
element of the ensemble from another. Each of the particles in the
$\al$-th system is in a quite definite elementary state
$[\vp^{\al}_{\xi 1}]$ for the first particle and in
$[\vp^{\al}_{\xi 2}]$ the second. In the case under consideration,
it is convenient to take the three-dimensional unit vector ${\bf
n}$ as the index  $\xi$  fixing the commutative subalgebra. In
this case, we must assume that the same subalgebra corresponds to
the vectors ${\bf n}$ and   ${\bf -n}$. Then the characters
$\vp^{\al}_{\xi 1}$ and $\vp^{\al}_{\xi 2}$  the elementary states
describe the (quasi)spin projections on the direction ${\bf n}$
for the first and the second particle. They take the values
\cc{slav1}): $\vp^{\al}_{{\bf n}i}=S^{\al}_{{\bf n}i} \qquad
(i=1,2)$, where  $S^{\al}_{{\bf n}i}=-S^{\al}_{{-\bf n}i}$, and
$S^{\al}_{{\bf n}1}=-S^{\al}_{{\bf n}2}$, and $S^{\al}_{{\bf n}1}$
is equal to either $+1/2$ or $-1/2$ depending on   $\al$ and ${\bf
n}$. Under such conditions, the elementary state of one particle
of the EPR pair is the negative copy of the elementary state of
the other particle. Therefore, a measurement of the value of an
observable of the first particle is automatically a measurement of
the corresponding observable of the second particle irrespective
of the location of this particle. Such measurement is said to be
indirect. In the case of such a measurement, the measured object
is not subjected to the action of the measuring device. Therefore,
its elementary state remains unchanged. The strict correlation
between the elementary states of the particles in the EPR pair is
its characteristic feature. This correlation develops in the
prearrangement of the EPR pair rather than in the subsequent
measurement.   This correlation is not a consequence of the
projection postulate; just the opposite, the statements in the
projection postulate are a consequence of such a correlation.
Similar correlations are also inherent in other entangled states.

\section{Some devices of quantum physics}

To understand the quantum teleportation procedure, it is necessary
to understand the operation principles of the devices used. We
give a very brief description of three such devices in this
section.

In the preceding section, the EPR paradox was studied using an
example of a system consisting of two particles with spins 1/2.
But it is technically very difficult to create the required EPR
pair for such a system. It is considerably simpler to obtain such
a pair optically. The corresponding device is called a type-II
parametric frequency down-converter. In the Russian literature,
this device is often called a source of spontaneous parametric
scattering (SPR source). A nonlinear optical crystal irradiated by
a UV laser is the basis of the SPR source. Laser photons are
scattered in the crystal. For such scattering, one photon at the
input as a rule yields one photon at the output. But one photon
sometimes generates two photons with much lower (many orders
lower) intensities. For SPRs of the second type, these photons
turn out to be polarized in two mutually orthogonal directions $H$
(horizontal) and  $V$ (vertical). We can choose conditions such
that these photons form an EPR pair, i.e., their quantum state can
be described by the vector  $|\Psi^{(-)}\ra_{12}$ (see
formula~\rr{3}) where $|+\ra$ and $|-\ra$ correspond to the
respective horizontal and vertical polarizations.

The second device is a (simple) beam splitter. It is designed to
mix two photon beams without changing their intrinsic
characteristics. We can visualize this device as a semitransparent
plate (see Fig. 1) with two photon beams are incident at equal
angles from above and below in one plane. If photons from
different beams are incident on the plate at different instants,
then each of them passes through the plate independently with
probability 1/2 or reflects from it with probability 1/2. But if
two photons are simultaneously incident, then they interfere
according to the formula
  \bea{6}
|H,V\ra_{u}^{in}&\to&\frac{1}{\sqrt{2}}\lt[|H,V\ra_u^{out}+
|H,V\ra_d^{out}\rt],\nn
|H,V\ra_{d}^{in}&\to&\frac{1}{\sqrt{2}}\lt[|H,V\ra_u^{out}-
|H,V\ra_d^{out}\rt],
 \eea
where the subscripts  $u$ and $d$ denote the upper and lower beams
and  $|H,V\ra$ means either  $|H\ra$ or $|V\ra$. For definiteness,
Fig. 1 shows the case where the upper incident beam is
horizontally polarized (it is described by the vector
$|H\ra_u^{in}$) and the lower beam is vertically polarized
$|V\ra_d^{in}$.

 \begin{figure}[h]
 \begin{center}

  \includegraphics{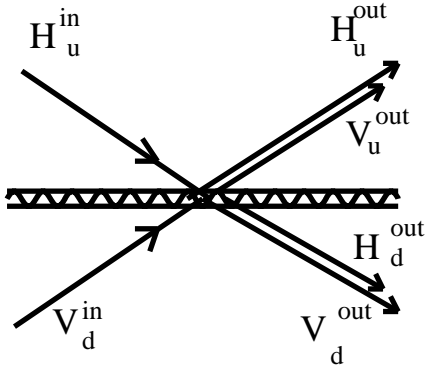}

   \caption{Simple beam splitter.}
\end{center}
\end{figure}

It follows from formulas  \rr{3} that as a result of passing
through the beam splitter, the Bell states (see formulas \rr{6})
transform as

  \bea{7}
|\Psi^{(-)}\ra_{ud}^{in}&\to&\frac{1}{\sqrt{2}}\lt[|H\ra_u^{out}|V\ra_d^{out}-
|V\ra_u^{out}|H\ra_2\rt]\equiv-|\Psi^{(-)}\ra_{ud}^{out},\nn
|\Psi^{(+)}\ra_{ud}^{in}&\to&\frac{1}{\sqrt{2}}\lt[|H\ra_u^{out}|V\ra_u^{out}-
|H\ra_d^{out}|V\ra_d^{out}\rt],\nn
|\Phi^{(-)}\ra_{ud}^{in}&\to&\frac{1}{\sqrt{2}}\lt[|H\ra_u^{out}|H\ra_u^{out}-
|V\ra_u|V\ra_u^{out}-|H\ra_d^{out}|H\ra_d^{out}+
|V\ra_d^{out}|V\ra_d^{out}\rt],\nn
|\Phi^{(+)}\ra_{ud}^{in}&\to&\frac{1}{\sqrt{2}}\lt[|H\ra_u^{out}|H\ra_u^{out}+
|V\ra_u^{out}|V\ra_u^{out}-|H\ra_d^{out}|H\ra_d^{out}-
|V\ra_d^{out}|V\ra_d^{out}\rt].
 \eea
In other words, after the beam splitter, the photons in each
simultaneous pair turn out to be on different sides of the plate
if they are incident in the state  $|\Psi^{(-)}\ra_{ud}^{in}$
(quasispin $0$) and on the same side for the states
$|\Psi^{(+)}\ra_{ud}^{in}, \quad|\Phi^{(-)}\ra_{ud}^{in}$, and
$\quad |\Phi^{(+)}\ra_{ud}^{in}$ (quasispin 1).

The third device is a polarization beam splitter (PBS). The
geometry of this device determines the so-called (orthogonal)
polarization basis. A photon beam incident on the PBS is directed
along the first basis vector. The other two basis vectors
determine the vertical and horizontal directions. The beam passes
through the PBS if it is polarized in the horizontal direction and
is reflected if it is polarized in the vertical direction. The PBS
is often used in combination with detectors located in the
propagation directions of the secondary beams. Using such a
device, we can determine photon polarization.

If the PBS is rotated through an angle about the first basis
vector, then we obtain a new polarization basis. After a beam with
certain polarization in the original basis passes through the PBS,
it decomposes into two subbeams each of which has a certain
(horizontal or vertical) polarization in the new basis. If the
detectors used in experiments are one-photon detectors (sensitive
to separate photons), then we can determine which polarization a
separate photon has in the new basis. According to the standard
ideology of quantum mechanics, a separate photon of the beam
polarized in the direction of one of the vectors of the original
basis has no definite polarization with respect to the vectors of
the rotated basis. Passing through the PBS, it acquires such a
polarization. In this case, such a selection process has no
objective reason. It is purely random in this sense.

According to the ideology in Sec. 2, each separate photon has a
particular (vertical or horizontal) polarization in any
polarization basis. All these polarizations depend on the
elementary state of this photon. But we initially know only the
polarization in the original basis. A rotated PBS classifies
photons of the original beam according to their polarizations in
its polarization basis. At the same time, the PBS perturbs the
previously known polarization in the original basis. Thus, using
PBSs and one-photon detectors, we can determine the polarization
of each photon in any direction but only one at a time.

\section{Quantum teleportation}

Figure 2 shows a scheme of quantum teleportation.

\begin{figure}[h]
 \begin{center}

  \includegraphics{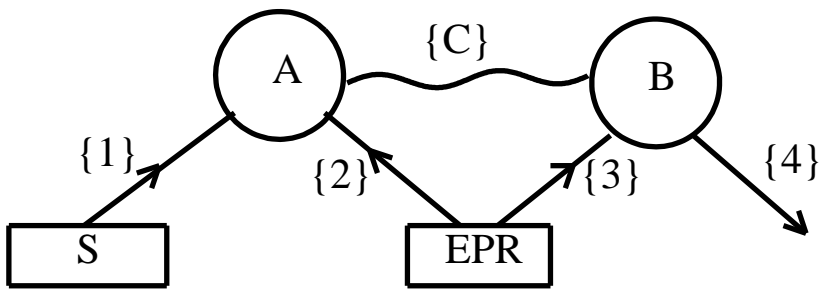}

   \caption{ Scheme of quantum teleportation.}
\end{center}
\end{figure}

Here, $S$ is the source of the initial state,  $EPR$ is the source
of EPR pairs, $A$ is the analyzer of the Bell states (Alice),  $B$
is the unitary converter (Bob),  $\{C\}$ is the classical
communication channel, \{1\} is the carrier of the initial
teleported state,  \{2\} -- \{3\} is the EPR pair, and \{4\} is
the carrier of the final teleported state.

We give the standard description of the teleportation scheme (see,
e.g.,~\cc{bwz}). Each of the particles  \{1\},\{2\},\{3\}, and
\{4\} participating in the process can be in one of the quantum
levels  $|+\ra$ or $|-\ra$. The source S emits particle \{1\} in
the quantum state  $|\Psi\ra_1=\al|+\ra+\beta|-\ra$, where
$|\al|^2+|\beta|^2=1$. In the general case, $\al$ and  $\beta$ can
be unknown. The EPR source emits particles  \{2\} and \{3\} in the
quantum state  $|\Psi^{(-)}\ra_{23}$ (see formula~\rr{3}). The
quantum state of the three-particle system consisting of particles
\{1\},\{2\}, and \{3\} is described by the vector
$|\Psi\ra_{123}=|\Psi\ra_1\bigotimes|\Psi^{(-)}\ra_{23}$ which can
be decomposed in terms of the Bell states of particles \{1\} and
\{2\}:
 \bea{8}
|\Psi\ra_{123}&=&\frac 12\lt\{|\Psi^{(-)}\ra_{12}\Big(-\al|+\ra_3
- \beta|-\ra_3\Big)+|\Psi^{(+)}\ra_{12}\Big(-\al|+\ra_3 +
\beta|-\ra_3\Big)\rt.\nn \
&+&\lt.|\Phi^{(-)}\ra_{12}\Big(\al|-\ra_3 +
\beta|+\ra_3\Big)+|\Phi^{(+)}\ra_{12}\Big(\al|-\ra_3 -
\beta|+\ra_3\Big)\rt\}.
 \eea
Using the analyzer  $A$, Alice measures to find the Bell states of
the four possible for the accessible particles \{1\} and \{2\}.
For example, we suppose that she obtains  $|\Psi^{(-)}\ra_{12}$ as
a result. Then the three-particle system collapses to the state
$|\Psi'\ra_{123}=|\Psi^{(-)}\ra_{12}\Big(-\al|+\ra_3 -
\beta|-\ra_3\Big)$ after such a measurement according to the
projection postulate. Alice broadcasts her discovery that the
particles \{1\} and \{2\} are in the state  $|\Psi^{(-)}\ra_{12}$
over the classical communication channel. Bob, doing nothing,
transmits particle  \{3\}. This particle is now in the state
$|\Psi\ra_4=\Big(-\al|+\ra_3 - \beta|-\ra_3\Big)$, which coincides
with the state  $|\Psi\ra_1$. If Alice finds another Bell state,
then she uses the classical communication channel to report to Bob
which of the three unitary transformations
($-|+\ra\longrightarrow|+\ra,\quad |+\ra\longleftrightarrow|-\ra,$
and $-|+\ra\longrightarrow|+\ra\longleftrightarrow |-\ra$) he must
apply to particle  \{3\} to obtain the state $|\Psi\ra_4$
coinciding with  $|\Psi\ra_1$.

The case where Alice discovers the result $|\Psi^{(-)}\ra_{12}$.
seems especially mysterious in such a description of the
teleportation process. There was no correlation between particles
\{1\} and \{3\} before the measurement. It is very difficult to
imagine how Alice, not acting physically on particle  \{3\}, could
make it pass to the quantum state of particle \{1\}. In this case,
Alice even knew nothing about that state.

We now discuss how the same teleportation process can be described
using the notion of an elementary state. In this case, a whole
beam of particles \{1\} that are in the different elementary
states (but all belonging to the same equivalence class)
corresponds to the quantum state  $|\Psi\ra_1$. Accordingly, a
beam of EPR pairs \{2\}-\{3\} rather than one pair is required in
the experiments. The numbers $\al$ and $\beta$ in the quantum
state  $|\Psi\ra_1$ specify a direction  ${\bf n}$, along which
the quasispin projection of each particle  \{1\} of the beam is
definitely equal to 1/2. Let the $z$ axis be along the direction
${\bf n}$. Then, for the quasispin projection, the equality
$S_z=+1/2$ holds in the quantum state  $-\al|+\ra_3 -
\beta|-\ra_3$, the equality $S_z=-1/2$ holds in the state
$-\al|+\ra_3 + \beta|-\ra_3$, the equality $S_x=+1/2$ holds in the
state $\al|-\ra_3 + \beta|+\ra_3$, and the equality $S_x=-1/2$
holds in the state $|\al|-\ra_3 - \beta|+\ra_3$.

We now regard the analyzer  $A$ in combination with particle \{1\}
as a measuring device. The action of this combined measuring
device on the beam of particles  \{2\} can be interpreted two ways
(see formula~\rr{8}). On one hand, this device divides the beam of
particles  \{2\} into four subbeams, in each of which particles
\{2\} (in combination with particles  \{1\}) are in one of the
Bell states. This result is fixed by the analyzer  $A$. On the
other hand, the particles  \{2\} in each of these four subbeams
have definite values of the quasispin projections either on the
$z$ axis or on the  $x$ axis. Because of the strict correlation
between the elementary states of particles  \{2\} and \{3\}, the
beam of particles  \{3\} automatically divides into four subbeams
in each of which particles \{3\} have certain quasispin
projections. That is, we have a typical example of an indirect
measurement of the quasispin projection for particle \{3\} in this
case. Using the classical communication channel, Alice reports the
result of such an indirect measurement to Bob. Bob applies the
corresponding unitary transformation to particles  \{3\}. As a
result of this measurement, only some information about this
elementary state needed for Bob's subsequent actions is obtained.

We call attention to the fact that the elementary state of
particle  \{3\} does not become the same as that of particle \{1\}
after all the described manipulations. These particles only turn
out to be in the same equivalence class. Thus, particle \{3\} does
not become an exact copy of particle  \{1\}; therefore, the term
"teleportation" used to describe this procedure is not especially
appropriate.

We now discuss an actual experiment in which teleportation was
observed~\cc{bpwz}. The schematic diagram of the experimental
setup is shown in Fig. 3.

\begin{figure}[h]
 \begin{center}

  \includegraphics{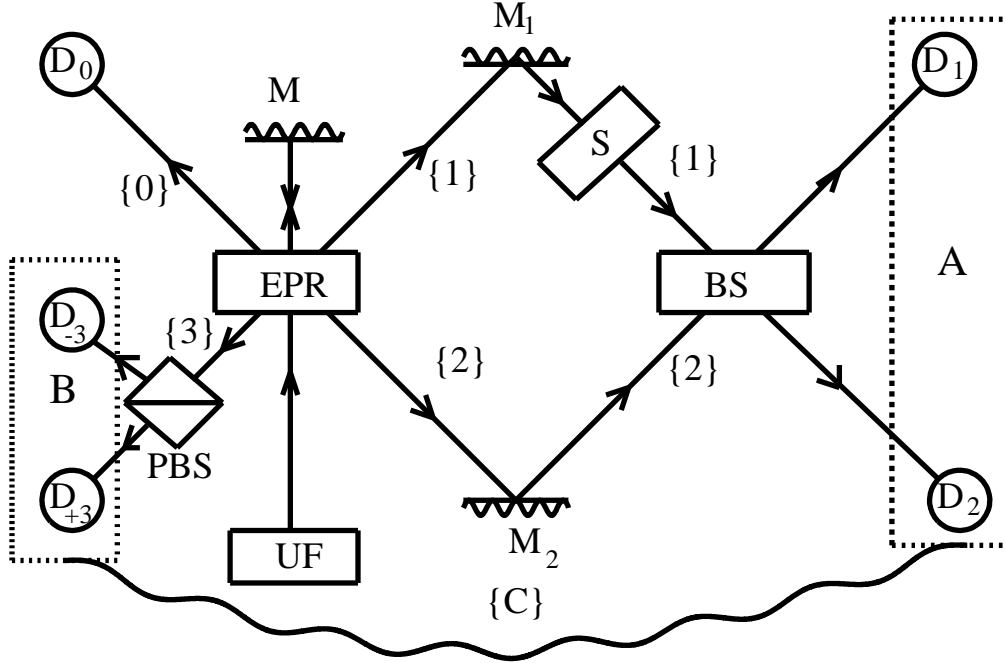}

 \caption{Schematic diagram of the quantum teleportation
experiment.}
\end{center}
\end{figure}

Here, $UF$ is the laser (the source of ultraviolet pulses),  $EPR$
is the EPR source;  $M$, $M_1$, and $M_2$ are totally reflecting
mirrors, $BS$ is the simple beam splitter,  $PBS$ is the
polarization beam splitter, $S$ is the coder of the initial state
of photon  \{1\}, $D_0$, $D_1$, $D_2$, $D_{+3}$, and $D_{-3}$ are
the detectors, and  \{C\} is the classical communication channel
(the coincidence circuit for detector combinations).

For simplicity of argument, we assume that all photons under
consideration propagate in the same plane and that the horizontal
direction of the original polarization basis is in this plane.

A laser UV pulse is incident on a nonlinear crystal (EPR). The EPR
pair  \{0\}-\{1\}. is produced in this crystal. After the pulse is
transmitted through the crystal, it is reflected from the mirror
$M$ and again enters the crystal, where it produces the second EPR
pair \{2\}-\{3\}. Photon \{1\} reflecting from the mirror  $M_1$,
enters the coder  $S$, where it acquires a particular
polarization. Two versions were studied in the actual experiment:
polarization at a  $45^0$ angle and at a  $90^0$ angle. After
this, photon \{1\} enters the beam splitter  $BS$.

Photon \{2\} reflecting from the mirror $M_2$ enters the same beam
splitter  $BS$ but on the other side. Moving the mirror $M$, we
can ensure that photons  \{1\} and \{2\} are inserted in the beam
splitter $BS$ simultaneously.  Then they interfere in the beam
splitter; after passing through it, either they both enter  one of
the detectors  $D_1$ or $D_2$ or one of them enters the detector
$D_1$ and the other enters  $D_2$.

Photon \{3\} is directed to the polarization beam splitter and
then enters either the detector $D_{+3}$ or $D_{-3}$ depending on
the polarization. Photon \{0\} is immediately directed to the
detector $D_0$. The coincidence circuit  \{C\} registers events in
which either the detectors  $D_0$, $D_1$, $D_2$, and $D_{+3}$ or
the detectors  $D_0$, $D_1$, $D_2$, and $D_{-3}$ operate
simultaneously.

It was taken into account in the experiment that the creation of
an EPR pair is a very rare event compared with the creation of
single photons, and the creation of two EPR pairs is a much rarer
event. Therefore, it is very important to separate random
coincidences in the detector operation from significant
coincidences. In this regard, the detector  $D_0$ is required to
separate the events containing photon \{1\}, whose state must be
teleported. The experimenters interpreted coincidence in the
operation of the detectors  $D_1$ and $D_2$ as evidence that
photons  \{1\} and \{2\} turn out to be in the quantum state
$\Psi^{(-)}\ra_{12}$ after such a measurement. Accordingly, photon
\{3\} acquires the same quantum state as photon \{1\} after
passing through the coder $S$. To verify this, the polarization
beam splitter  $PBS$ was oriented such that if photon \{3\} had
the same polarization as encoded in photon \{1\}, then photon
\{3\} must enter the detector  $D_{+3}$. Therefore, the situation
in which the coincidence $D_0-D_1-D_2-D_{+3}$ is observed and
there is no coincidence  $D_0-D_1-D_2-D_{-3}$ must correspond to
the case of teleportation.

Actually, the experimenters studied the rate of coincidences of
such type depending on the position of the mirror  $M$, i.e.,
depending on how photons  \{1\} and \{2\} simultaneously enter the
beam splitter  $BS$, in other words, whether they interfere in the
beam splitter or are independently scattered. For total
interference, the number of coincidences $D_0-D_1-D_2-D_{-3}$ must
decrease to zero. If there is no interference, both types of
coincidences turn out to be random, and their rates become the
same. Indeed, a considerable dip (of approximately one order)
corresponding to the interference of photons \{1\} and \{2\} was
observed in the graph of the dependence of the rate of
coincidences $D_0-D_1-D_2-D_{-3}$ on the position of the mirror
$M$. This was confirmed in subsequent similar experiments
\cc{jwpz}.

But a new result not easily interpreted was also obtained. The
point is that by changing the lengths of the paths of photons
\{1\} and \{2\}, we can ensure that the described effect is
attained in the case where the time delay for the detectors $D_1$
and $D_2$ is inserted in the coincidence circuit. In this case,
the teleportation registered by the detectors  $D_{+3}$ and
$D_{-3}$ is obtained earlier than that registered by the detectors
D1 and D2. Such a result was indeed observed. Moreover, we can in
principle ensure that the teleportation is fixed before photon
\{1\} acquires a particular polarization. Strictly speaking, all
this does not formally contradict the standard mathematical
apparatus of quantum mechanics, but it is hard to believe that the
future can affect the past.

We now consider how the same experiment can be interpreted in the
framework of the scheme described in Sec. 2. We first consider the
case where photon  \{1\} passing through the coder is vertically
polarized in the original basis. Let the polarization beam
splitter $PBS$ be rotated through a $90^0$ angle with respect to
the original basis. Then photon  \{3\} enters the detector
$D_{+3}$ if it is polarized vertically in the original basis but
enters the detector $D_{-3}$ if it is horizontally polarized.

We consider the case where the detector  $D_{-3}$  operates.
Because photons \{2\} and \{3\} form an EPR pair, photon  \{2\} is
vertically polarized in this case. Hence, both photons  \{1\} and
\{2\} entering the beam splitter  $BS$ have the same polarization
in this case. Therefore, if they turn out to be in the beam
splitter  $BS$ simultaneously, then they both enter the same
detector (either $D_1$ or $D_2$) after the beam splitter (see
formula~\rr{7}). That is, the probability of the coincidence
$D_0-D_1-D_2-D_{-3}$ is theoretically zero. In fact, this
probability is nonzero at least because the EPR source does not
produce ideal pairs.

In contrast to the standard quantum mechanics, a strict time
correlation between the events is predicted in the given scheme.
Within a small error related to the duration of the laser pulse
and to the fact that photons \{2\} and \{3\}  can be emitted at
different instants, the positive effect can be reached if the
coincidence circuit is adjusted as follows. The instants  $t_0$
(the time when photon  \{0\} attains the detector  $D_0$),
$t_{12}$ (the time when photons \{1\} and \{2\} attain the
detectors  $D_1$ and $D_2$) and $t_3$ (the time when photon  \{3\}
attain the detector $D_{-3}$ or $D_{+3}$) must be assumed to be
simultaneous for it. The times $t_0$, $t_{12}$, and $t_3$ can be
in any ratio astronomically.

We now consider the case where photon \{1\} is polarized at a
$45^0$ angle in the original polarization basis. Let the detector
$D_{-3}$ register a photon polarized at a  $-45^0$ angle. Then
photon \{2\} is polarized at a  $45^0$  angle. The subsequent
discussion is easier in terms of quasispin than in terms of
polarization. We associate the quasispin orientation along the $x$
axis with the polarization at $45^0$ , i.e., $|45^0\ra\to
|S_x=1/2\ra$. Hence, in the case under consideration, the
elementary state of two-particle system \{12\} consisting of
photons \{1\} and \{2\} belongs to the quantum state
$|S_x=1/2\ra_1|S_x=1/2\ra_2$. According to the standard rules of
quantum mechanics, the total quasispin is equal to 1 in this
quantum state. Therefore, according to formulas~\rr{7} both
photons passing through the beam splitter $BS$ must enter the same
detector  ($D_1$ or $D_2$). As a result, the coincidence circuit
does not register this case, i.e., the result is the same as for
the photon polarization at a  $90^0$ angle.

But the situation can turn out to be more complicated in
actuality. The point is that two-particle system  \{12\} was not
prearranged as a quantum system with the value of the squared
total quasispin  $S^2$ equal to 2. The elementary states of
two-particle system  \{12\} were not selected using the value of
the observable $\s^2$. Therefore, there is no guarantee what such
a value will necessarily be in each elementary event. At best, if
we can believe Postulate 7, we can only state that the mean of the
observable  $\s^2$ in an infinite number of trials is equal to 2.

In the actual experiment, the number of cases where the presence
of coincidences  $D_0-D_1-D_2-D_{+3}$ and the absence of
coincidences $D_0-D_1-D_2-D_{-3}$ are registered is very small.
Therefore, in each trial, the probability of the deviation of the
value of the observable  $\s^2$ from the mean is sufficiently
large. This probability must decrease to improve the statistics.
It seems to follow from the preceding that it is useless to
improve the statistics in an attempt to discover the deviation
from the predictions of standard quantum mechanics. But this is
not so. We must only treat these deviations as the effect under
study rather than as measurement errors. For example, we can study
the quantity
   $$     \rho=\frac{N_-(45)}{N_-(90)} \qquad \mbox{при}\quad
   N_-(45)+N_+(45)=N_-(90)+N_+(90),   $$
where $N_-(45)$ is the number of coincidences $D_0-D_1-D_2-D_{-3}$
in the absence of coincidences  $D_0-D_1-D_2-D_{+3}$, $N_+(45)$ is
the number of coincidences $D_0-D_1-D_2-D_{+3}$ in the absence of
coincidences  $D_0-D_1-D_2-D_{-3}$ for the polarization of photon
\{1\} at a $45^0$ angle, and $N_-(90)$ and $N_+(90)$ are the same
numbers for the polarization at a $90^0$ angle. If the standard
quantum mechanics is used in the case under consideration, then it
must be expected that the value of $\rho$ must approach unity to
improve the statistics. But if we use the approach proposed here,
then it must be expected that the value of $\rho$ is always
greater than unity. We note that a similar result was already
discovered experimentally~\cc{jwpz}). But the experimenters were
prone to think that this result was an experimental error. In such
a case, a special experiment should be conducted to study this
effect more thoroughly.

\section{Teleportation of entanglement}

The phenomenon that obtained the name "teleportation of
entanglement" is often considered the strongest argument in favor
of nonlocality inherent in quantum measurements. This phenomena
can be understood from Fig. 4, where $EPR\;1$ and $EPR\;2$ are the
sources of independent EPR pairs, $A$ is the analyzer of the Bell
states (Alice), $B$ is the unitary converter (Bob),  \{C\} is the
classical communication channel (the coincidence circuit),
\{0\}-\{1\} is the first EPR pair, and  \{2\}-\{3\} is the second
EPR pair.

\begin{figure}[h]
 \begin{center}

  \includegraphics{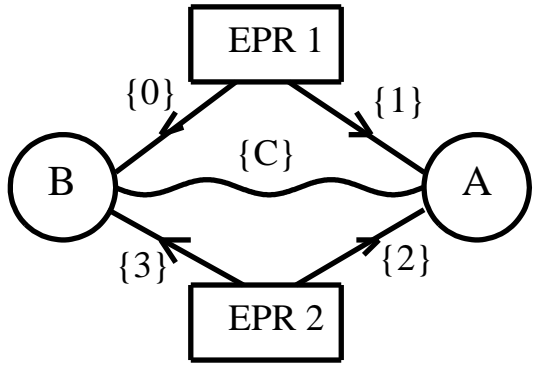}

   \caption{Scheme for teleportation of entanglement.}
\end{center}
\end{figure}

As before, we first give the standard description of the scheme
for teleportation of entanglement~\cc{bpwz1}. The source $EPR\;1$
emits the pair  \{0\}-\{1\} in the quantum state
$|\Psi^{(-)}\ra_{01}$. The source  $EPR\;2$ independently emits
the pair \{2\}-\{3\} in the quantum state $|\Psi^{(-)}\ra_{23}$.
The vector of the four-particle quantum state
$|\Psi\ra_{0123}=|\Psi^{(-)}\ra_{01}\bigotimes|\Psi^{(-)}\ra_{23}$
can be decomposed in terms of the Bell states of particles
\{1\}-\{2\}:
 $$
|\Psi\ra_{0123}=\frac 12 \lt\{
|\Psi^{(+)}\ra_{03}|\Psi^{(+)}\ra_{12}-
|\Psi^{(-)}\ra_{03}|\Psi^{(-)}\ra_{12}-
|\Phi^{(+)}\ra_{03}|\Phi^{(+)}\ra_{12}+
|\Phi^{(-)}\ra_{03}|\Psi^{(-)}\ra_{12}\rt\}.
  $$
This formula is simultaneously a decomposition of the vector
$|\Psi\ra_{0123}$ in terms of the Bell states of particles
\{0\}-\{3\}. Using the analyzer  $A$, Alice determines the Bell
state from the four possibilities for the pair  \{1\}-\{2\} and
reports the result to Bob using the classical communication
channel. For example, let  $|\Psi^{(-)}\ra_{12}$ be such a state.
According to the projection postulate, the state $|\Psi\ra_{0123}$
then collapses to the state
$|\Psi^{(-)}\ra_{03}|\Psi^{(-)}\ra_{12}$. Therefore, Bob can now
state that the pair  \{0\}-\{3\} is in the quantum entangled state
$|\Psi^{(-)}\ra_{03}$.

Particles \{0\} and \{3\} were initially completely independent.
Further, they were not subjected to any physical action. All
manipulations were performed with particles  \{1\} and \{2\}.
Nevertheless, after these manipulations, particles  \{0\} and
\{3\} somehow mysteriously turned out to be in the entangled
state. We can give a much more obvious description of the
correlation between particles \{0\} and \{3\} in the framework of
the approach assumed here.

Using the analyzer  $A$,  Alice divides the beam of particles
\{1\} and \{2\} into four subbeams by using a criterion for the
presence of a particular correlation between the particles in each
subbeam. These correlations are not produced by the analyzer $A$;
the latter only serves to select pairs in which such correlations
appear randomly when these particles were emitted by the sources
$EPR\;1$ and $EPR\;2$.

The elementary state of particle  \{0\} is strictly correlated
with the elementary state of particle  \{1\}, and there is a
similar case for particles  \{2\} and \{3\}. Therefore, dividing
the beam of particles  \{1\} and \{2\} into subbeams automatically
divides the beam of particles  \{0\} and \{3\} into the
corresponding subbeams. In each pair  \{0\}-\{3\}, there is a
correlation that is the (negative) copy of the correlation in the
pair \{1\}-\{2\}. Using the classical communication channel \{C\},
we can establish which subbeam the particular pair  \{0\}-\{3\}
belongs to. Here, we must keep in mind that the correlation in the
pair \{0\}-\{3\} is the copy of the correlation in the pair
\{1\}-\{2\} that existed before the particles passed through the
analyzer  $A$. At the same time, the analyzer $A$ can perform the
function of not only a measuring device but also a device
prearranging a new elementary state. Therefore, the correlation in
the pair  \{0\}-\{3\} is generally not an exact copy of the
correlation in the pair \{1\}-\{2\} after the pair passes through
the analyzer  $A$.

Further, we discuss actual experiments in which the phenomenon of
teleportation of entanglement was studied~\cc{bpwz1,jwpz}. The
schematic diagram of the experimental setup is shown in Fig. 5.
For the most part, it is similar to the experimental setup shown
in Fig. 3. The difference is that there is no coder of the initial
state in Fig. 5; on the other hand, there are two beam splitters
$PBS_0$ and $PBS_3$  instead of one polarization beam splitter
$PBS$ and  two detectors $D_{-0}$ and $D_{+0}$  instead of one
detector $D_0$.

\begin{figure}[h]
 \begin{center}

  \includegraphics{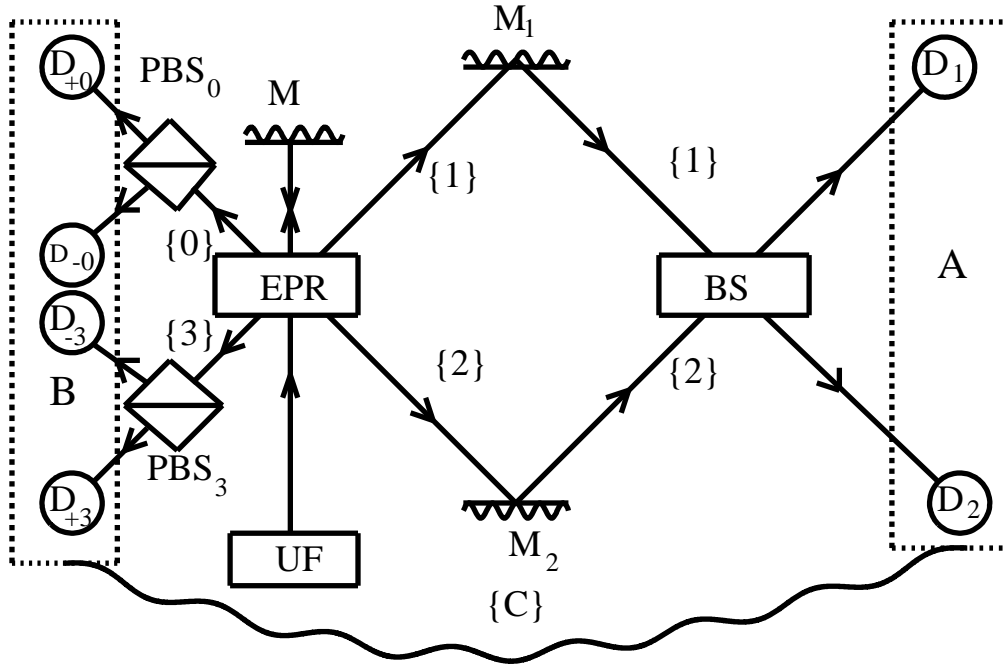}

   \caption{Schematic diagram for the experiment on teleportation of entanglement.}
\end{center}
\end{figure}

The detector  $D_{+0}$ operates when photon   \{0\} is
horizontally polarized in the polarization basis of the beam
splitter $PBS_0$. The detector  $D_{-0}$ operates when photon
\{0\} is vertically polarized. The detectors  $D_{+3}$ and
$D_{-3}$ operate similarly, but their response is connected with
the polarization basis of the beam splitter  $PBS_3$.

The experimenters studied the versions of the coincidences
$D_1-D_2-D_{\pm 0}-D_{\pm 3}$ for different orientations of the
polarization beam splitters with respect to the original
polarization basis. They assumed that in the case where the
detectors $D_1$ and $D_2$ operate simultaneously, photons \{1\}
and \{2\} turn out to be in the entangled two-particle state
$|\Psi^{(-)}\ra_{12}$. Therefore, the quantum two-particle state
of photons  \{0\} and \{3\} is  $|\Psi^{(-)}\ra_{03}$ according to
the projection postulate. Hence, the entangled state of photons
\{1\} and \{2\} produced using the beam splitter $BS$ and
detectors $D_1$ and $D_2$ is transferred to photons  \{0\} and
\{3\} without any physical action. These photons were completely
independent before the manipulations with photons  \{1\} and
\{2\}.

The following facts are considered to prove that photons  \{0\}
and \{3\} really turn out to be in the quantum state
$|\Psi^{(-)}\ra_{03}$. The fourfold coincidences $D_1-D_2-D_{+
0}-D_{- 3}$ and $D_1-D_2-D_{- 0}-D_{+ 3}$ reach their maximums and
the coincidences $D_1-D_2-D_{+ 0}-D_{+ 3}$ and $D_1-D_2-D_{-
0}-D_{- 3}$ reach their minimums for the same orientation of the
polarization beam splitters $PBS_0$ and $PBS_3$. Conversely, if
the beam splitters  $PBS_0$ and $PBS_3$ are orthogonal to each
other, then the coincidences  $D_1-D_2-D_{+ 0}-D_{+ 3}$ and
$D_1-D_2-D_{- 0}-D_{- 3}$ reach their maximums and the
coincidences  $D_1-D_2-D_{+ 0}-D_{- 3}$ and $D_1-D_2-D_{- 0}-D_{+
3}$. reach their minimums. Under simultaneous rotation of the
polarization beam splitters through the same angle, the maximums
remain maximums and the minimums remain minimums. This fact was
interpreted as evidence of the spherical symmetry of the
two-particle quantum state of photons \{0\} and \{3\}, which is
typical of the state $|\Psi^{(-)}\ra_{03}$.

But the actual experiment did not show complete spherical
symmetry. The point is that the depths of the maximums and
minimums change under the described rotation of the polarization
beam splitter axes, reaching their largest values when the axes of
the polarization beam splitters  $PBS_0$ and $PBS_3$ are rotated
through either $0^0$ or  $90^0$ with respect to the original
polarization basis and their least values for  $45^0$.

In the framework of the approach proposed here, the results of
similar experiments can be interpreted by a method similar to that
described in Sec. 5. The simple beam splitter  $BS$ and the
detectors $D_1$ and $D_2$ select those pairs from the set of
different pairs of photons  \{1\} and \{2\} whose photons have a
mutually orthogonal polarization in the original polarization
basis. In addition, the mean of the total quasispin of these pairs
is zero. Because the elementary state of photon  \{0\} is the
(negative) copy of the elementary state of photon \{1\}  and the
elementary state of photon  \{2\} is the (negative) copy of the
elementary state of photon \{3\}, the pairs of photons  \{0\} and
\{3\} with the same properties are simultaneously selected.
Therefore, in the case where the polarization beam splitters
$PBS_0$ and $PBS_3$ are oriented at an angle of  $0^0$ or $90^0$
with respect to the original basis, the polarizations of photons
\{0\} and \{3\} must be orthogonal to each other in the
polarization bases related to the beam splitters  $PBS_0$ and
$PBS_3$. Accordingly, the versions of the coincidences observed in
the experiment must indeed be observed. In the case where the
polarization beam splitters  $PBS_0$ and $PBS_3$ are oriented at a
$45^0$ angle with respect to the original basis, we can advance
the same arguments used in Sec. 5. As a result, we conclude that
for an infinite number of trials, the polarizations of photons
\{0\} and \{3\} measured using the beam splitters  $PBS_0$ and
$PBS_3$ are orthogonal on the average. For a small number of
trials, we can observe considerable deviations from this rule.

Thus, in the experiment described above, we have the selection of
events involving the approximate desired correlation and not
teleportation of entanglement.

\section{Conclusions}

There is a widespread opinion that teleportation is a special
quantum method for transferring information and that this method
can turn out to be very effective. But it follows from our
considerations in this paper that there need not be a close
relation between teleportation and the quantum features. We can
easily give a classical analogue of quantum teleportation that has
long been used in practice. Before setting out to sea, the captain
of a ship obtains a sealed envelope from his headquarters
containing several numbered variants of the instruction
prescribing subsequent actions. When the ship is at sea, the
captain receives an order by radio to open the envelope and act
according to the instruction bearing a certain number. In quantum
teleportation, a particle from the EPR pair plays the role of the
envelope, and a classical communication channel plays the role of
the radio.

Teleportation changes the traditional method of the action
slightly. Instead of one envelope with an instruction, an
automatic device prepares two identical envelopes with
instructions. In these instructions, the action variants are
numbered identically but randomly. One envelope is sent to the
captain on the ship, and the other is sent to his headquarters on
shore. Before the envelopes are opened, nobody knows the number
assigned to each variant. Further, the chief opens his envelope,
chooses the necessary variant, and sends its number to the
captain. Such a method of action has a higher degree of secrecy
compared with the traditional one. But a higher price must be paid
for this advantage. In the case of the traditional method of
action, the envelope could be delivered to the captain before the
departure. In the case of quantum teleportation, the "envelope" is
sent while the captain is at sea. This can turn out to be a very
complicated technical problem. Therefore, one should not cherish
vain hopes to use the potential of quantum teleportation.

\end{document}